\documentclass{jfm}

\usepackage{graphicx}
\usepackage{newtxtext}
\usepackage{newtxmath}
\usepackage{natbib}
\usepackage{hyperref}
\hypersetup{
    colorlinks = true,
    urlcolor   = blue,
    citecolor  = black,
}

\newcommand{\RomanNumeralCaps}[1]
\linenumbers

\usepackage{setspace}
\usepackage{afterpage}
\usepackage[per-mode=symbol]{siunitx}
\usepackage{tabulary,graphicx,amsfonts,amsmath,amssymb,amsbsy,dcolumn,bm}
\usepackage[utf8]{inputenc}
\usepackage[T1]{fontenc}
\usepackage{gensymb}
\usepackage{graphicx}
\usepackage{dcolumn}
\usepackage{bm}
\usepackage{microtype}

\usepackage{xcolor}
\usepackage{hyperref}
\usepackage{tabularx} 
\usepackage{booktabs}
\usepackage{widetable}
\usepackage[final]{changes}
\usepackage{float}
\usepackage{siunitx}
\usepackage{amsmath}

\newlength{\toprulewidth}
\patchcmd{\toprule}
  {\heavyrulewidth}{\toprulewidth}
  {}{}

\hypersetup{
  colorlinks,
  citecolor=blue,
  linkcolor=red,
  urlcolor=blue}

\usepackage{url,multirow,morefloats,floatflt,cancel,tfrupee}
\makeatletter

\AtBeginDocument{\@ifpackageloaded{textcomp}{}{\usepackage{textcomp}}}
\makeatother
\usepackage{colortbl}
\usepackage{xcolor}
\usepackage{pifont}
\usepackage[nointegrals]{wasysym}
\makeatletter

\def\mcWidth#1{\csname TY@F#1\endcsname+\tabcolsep}

\def\cAlignHack{\rightskip\@flushglue\leftskip\@flushglue\parindent\z@\parfillskip\z@skip}
\def\rAlignHack{\rightskip\z@skip\leftskip\@flushglue \parindent\z@\parfillskip\z@skip}

\@ifundefined{etal}{}{}

\usepackage{ifxetex}
\ifxetex\else\if@twocolumn\@ifpackageloaded{stfloats}{}{\usepackage{dblfloatfix}}\fi\fi

\AtBeginDocument{
\expandafter\ifx\csname eqalign\endcsname\relax
\def\eqalign#1{\null\vcenter{\def\\{\cr}\openup\jot\m@th
  \ialign{\strut$\displaystyle{##}$\hfil&$\displaystyle{{}##}$\hfil
      \crcr#1\crcr}}\,}
\fi
}

\AtBeginDocument{%
  \@ifpackageloaded{endfloat}%
   {\renewcommand\efloat@iwrite[1]{\immediate\expandafter\protected@write\csname efloat@post#1\endcsname{}}}{\newif\ifefloat@tables}%
}%

\def\BreakURLText#1{\@tfor\brk@tempa:=#1\do{\brk@tempa\hskip0pt}}
\let\lt=<
\let\gt=>
\def\processVert{\ifmmode|\else\textbar\fi}

\@ifundefined{subparagraph}{
\def\subparagraph{\@startsection{paragraph}{5}{2\parindent}{0ex plus 0.1ex minus 0.1ex}%
{0ex}{\normalfont\small\itshape}}%
}{}

\newcommand\role[1]{\unskip}
\newcommand\aucollab[1]{\unskip}
  
\@ifundefined{tsGraphicsScaleX}{\gdef\tsGraphicsScaleX{1}}{}
\@ifundefined{tsGraphicsScaleY}{\gdef\tsGraphicsScaleY{.9}}{}
\def\checkGraphicsWidth{\ifdim\Gin@nat@width>\linewidth
	\tsGraphicsScaleX\linewidth\else\Gin@nat@width\fi}

\def\checkGraphicsHeight{\ifdim\Gin@nat@height>.9\textheight
	\tsGraphicsScaleY\textheight\else\Gin@nat@height\fi}

\def\fixFloatSize#1{}

\DeclareMathAlphabet{\mathpzc}{OT1}{pzc}{m}{it}

\def\URL#1#2{\@ifundefined{href}{#2}{\href{#1}{#2}}}

\def\UrlOrds{\do\*\do\-\do\~\do\'\do\"\do\-}%
\g@addto@macro{\UrlBreaks}{\UrlOrds}

\edef\fntEncoding{\f@encoding}

\makeatother

\newif\ifmultipleabstract\multipleabstractfalse%
\newcommand{\Ubold}{\textbf{u}}
\newcommand{\del}[2]{\frac{\partial#1}{\partial#2}}

\title{Laser-driven droplet deformation at low Weber numbers
}

\affiliation{\aff{1}Advanced Research Center for Nanolithography (ARCNL),\\Science Park 106, 1098 XG Amsterdam, The Netherlands
\aff{2}LaserLab, Department of Physics and Astronomy, Vrije Universiteit Amsterdam,\\ De Boelelaan 1100, 1081 HZ Amsterdam, The Netherlands
\aff{3}Van der Waals–Zeeman Institute, Institute of Physics, University of Amsterdam, 1098XH Amsterdam, The Netherlands
\aff{4}Fluids and Flows group, Department of Applied Physics,Eindhoven University of Technology, Eindhoven, the Netherlands}

\author{M. Kharbedia\aff{1}
\mbox{H. Franca}\aff{1,3}
\mbox{H.K. Schubert}\aff{1,2}
\mbox{D.J. Engels}\aff{1,2}
\mbox{M. Jalaal}\aff{3}
\mbox{O. O. Versolato}\aff{1,2}
\corresp{\email{versolato@arcnl.nl}}
}

\date{\today}
\begin{document}
\maketitle
\begin{abstract}

We investigate droplet deformation following laser-pulse impact at low Weber numbers ($\textrm{We}\sim0.1-100$). 
Droplet dynamics can be characterized by two key parameters: the impact We number and the width, W, of the distribution of the impact force over the droplet surface.
By varying laser pulse energy, our experiments traverse a phase space comprising (I) droplet oscillation, (II) breakup or (III) sheet formation.
Numerical simulations complement the experiments by determining the pressure width and by allowing We and W to be varied independently, despite their correlation in the experiments.
A single phase diagram, integrating observations from both experiments and simulations, demonstrates that all phenomena can be explained by a single parameter: the deformation Weber number $\textrm{We}_\textrm{d}=f(\textrm{We},\textrm{W})$ that is based on the initial radial expansion speed of the droplet, following impact.
The resulting phase diagram separates (I) droplet oscillation for $\textrm{We}_\textrm{d}<5$, from (II) breakup for $5<\textrm{We}_\textrm{d}<60$, and (III) sheet formation for $\textrm{We}_\textrm{d}>60$.
\\
\\
\textbf{Key words:} \textbf{Low-energy | Deformation | Oscillation | Breakup}

\end{abstract}\def\keywordstitle{Keywords} 

\maketitle

\section{Introduction}\label{sec:intro}

Upon impacting a solid surface, the momentum of the falling droplet is counterbalanced by the inertia, viscous dissipation, and surface tension, which governs fluid retraction (\cite{pasandideh1996}, \cite{josserand2016}). 
The resulting deformation exhibits complex dynamics, including oscillations, sheet formation during spreading, and jetting upon retraction (\cite{yarin_impact_1995}, \cite{zhang2022drop_forces_impact}). 
Similarly rich dynamics is observed when droplets splash on pillars (\cite{villermaux_drop_2011},  \cite{wang_bourouiba_2018_rim}). 
Deformations may also occur under the influence of external flows where a wide range of phenomena from droplet vibration (\cite{pilch1987use}, \cite{hsiang1992near}) to bag breakup (\cite{guildenbecher2009secondary}, \cite{Jalaal2012}, \cite{jackiw2021aerodynamic}) to sheet stripping (\cite{theofanous2011aerobreakup}, \cite{jalaal2014transient}) and catastrophic fragmentation (\cite{guildenbecher2009secondary}, \cite{theofanous2011aerobreakup}) may occur.
The resulting dynamics is often characterized by the Weber number that, $\textrm{We}=\rho D_0 U^2/\sigma$  (with density $\rho$, droplet diameter $D_0$, impact velocity $U$, and surface tension $\sigma$),  
and the Ohnesorge number that relates the viscous forces to inertial and surface tension forces. 
In contrast to previous methods, using a laser impact on a droplet offers a unique opportunity to precisely control the pressure profile on its surface. In wind tunnel experiments, the entire surrounding medium is in motion, which complicates any effort to stabilize or adjust the pressure profile on the droplet surface. Similarly, when employing a fixed pillar to perturb the droplet, the pillar remains in continuous contact with the droplet, and neither the tunnel nor the pillar techniques provides instantaneous control. As a result, in all these approaches, the droplet’s response is primarily determined by the impact velocity $U$.

\begin{figure}
    \centering
    \includegraphics[width=0.9\linewidth]{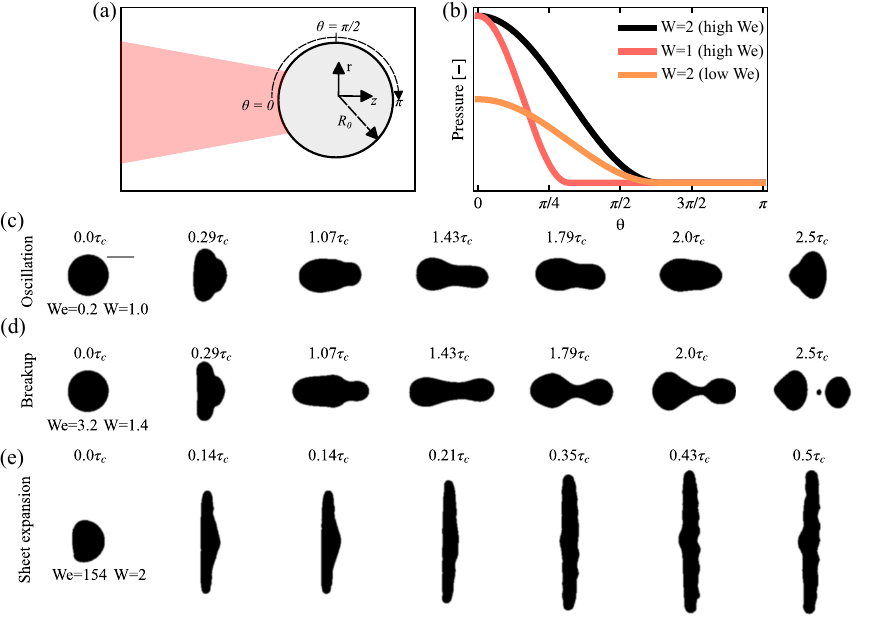}
    \caption{(a) Conceptual side view representation of laser-droplet interaction. 
    The laser beam is represented as a red area. 
    The force profile resulting from laser-plasma generation is depicted in (b) where curves show different values of pressure distribution on the droplet's surface. 
    The black and orange curves with the same value of $\textrm{W}$ depict different values of $\textrm{We}$ (higher and lower, respectively). 
    Experimental examples of the hydrodynamic response after laser interaction with a droplet with diameter $D_0=50\,\mu\textrm{m}$ is shown in (c-e). 
    Each row contains frames at different fractions of capillary times, $\tau_\textrm{c} = 16.4\,\mu\textrm{s}$.  From top to bottom: (c) droplet oscillation for $\textrm{W}=0.2$ and $\textrm{W}=1.0$.
    (d) droplet breakup after retraction for $\textrm{We}=3.4$ and $\textrm{W}=1.4$. 
    (e) sheet expansion for $\textrm{We}=154$ and $\textrm{W}=2$. The gray scalar bar in the first frame in (c) corresponds to $D_0=50\mu\mathrm{m}$. }
    \label{fig:1}
\end{figure}

Nevertheless, the force distribution on the surface is well known to significantly influence droplet deformation behavior at high We numbers (\cite{gelderblom_drop_2016}, \cite{hernandez-rueda_early-time_2022}, \cite{francca2025laser}), as established in the context of laser-driven droplet deformation for nanolithography (\cite{versolato_physics_2019}). 
The interaction between nanosecond-pulsed laser light with micrometer-sized tin droplets is a relevant source of extreme ultraviolet (EUV) light, as used in state-of-the-art industrial nanolithography.
The generation of EUV light involves the laser-impact deformation of a tin droplet into a thin sheet that is subsequently laser-heated into an EUV-emitting plasma. 
Upon ($\sim10\,\textrm{mJ}$) laser pulse impact, a plasma is generated from the droplet surface and produces a recoil pressure on the order of $\sim 100$\,kbar. 
In a typical setting, the droplet is accelerated on the laser pulse length scale, $\tau_{\textrm{p}}\sim10\textrm{ns}$, reaching terminal velocities of the order of $U\sim100\,\textrm{m/s}$ while it radially expands. 
The corresponding large Weber number ranges $\sim 1000-10\,000$, hence, the droplet transforms into a time-varying radially expanding liquid sheet (\cite{kurilovich_plasma_2016, kurilovich_power-law_2018}, \cite{liu_laser-induced_2021, liu_mass_2023}), undergoing several hydrodynamical instabilities responsible for its rupture through hole opening (\cite{klein_drop_2020}) or radial accumulation of the liquid into a fragmenting bounding rim (\cite{wang_bourouiba_2018_rim,wang_bourouiba_2021_growth}). 
The deformation has been shown to be well described by the impact We number and a dimensionless pressure width (\cite{gelderblom_drop_2016}, \cite{hernandez-rueda_early-time_2022}, \cite{francca2025laser}).
More specifically, \cite{francca2025laser} described the pressure profile in terms of a raised cosine function projected on the surface $\sim \left(1+\cos\left(\theta\pi/\textrm{W}\right) \right)H(\textrm{W}-\theta)$ [cf. fig.\,\ref{fig:1}(a,b)], where $H(\textrm{W}-\theta)$ is the Heaviside step function. This pressure profile accurately describes the sheet morphology over a range of pressure widths 1 < W < 2.5. 
Thus, unlike droplet impact on solid surfaces, falling drop, or wind tunnel interactions, laser-droplet interaction offers a path to studying how the spatial distribution of the force, along with its overall magnitude, influences droplet dynamics. 
However, motivated by direct nanolithography applications, prior studies have focused on high We numbers that lead to sheet formation.

In this work, we investigate droplet deformation following laser-pulse impact at low We numbers ($\textrm{We}\sim0.1-100$) combining experiments with numerical simulations. 
Decreasing drastically the pulse energy ($\sim0.1-1\,\textrm{mJ}$) in the experiment, we mainly focus on the dynamics of the regimes before sheet formation. 
In figs.\ref{fig:1}(c-e) we illustrate representative cases of droplet deformation over fractions of the capillary time $\tau_\textrm{c}=\sqrt{\rho D_0^3/6\sigma}$, with $\rho$ and $\sigma$ being liquid density and surface tension, respectively. 
At lower We numbers, an oscillatory motion is observed, as shown in fig.\,\ref{fig:1}(c), 
reminiscent of the previous work on oscillating free droplets (see e.g.,\cite{becker1991}, \cite{hsiang1992}, \cite{parik2024}, \cite{bostwick_capillary_2009_oscillations} and \cite{rimbert2020spheroidal}). 
The symmetry of this oscillation is unavoidably broken given the laser impact geometry. 
For slightly higher pulse energies the droplet breaks up, as shown in fig.\,\ref{fig:1}(d).
Finally, at even higher pulse energies, the radial inertia leads to sheet formation, as shown in fig.\,\ref{fig:1}(e). 
For reference, we provide estimates the corresponding (We, W) numbers for each case in fig.\,\ref{fig:1}, with W obtained using a scaling relation that also involves input from our simulations (see below).
The two parameters are strongly correlated in the experiment, with increased pulse energy leading to both increased We and W values.
The goal of this study is to characterize the individual influence of these two key parameters on the deformation process for low We numbers.
Our experiments traverse a phase space comprising (I) droplet oscillation, (II) breakup, a and (III) sheet formation.
Numerical simulations complement the experiment and allow for independently varying We and W.
 

This article is organized as follows. 
In \S\,\ref{sec:experimental} we provide a brief explanation of our experimental setup. 
Section \S\,\ref{sec:results} is devoted to the presentation and discussion of the results. 
The numerical simulations are explained in \S\,\ref{sec:CFD}. 
Next, in \S\,\ref{sec:scaling} we perform an analysis of the pressure profile on the droplet, where we explain how to correlate droplet's inertial deformation with the corresponding pressure projection on the surface. 
The oscillatory motion of the droplet is discussed in \S\,\ref{sec:oscillation}. 
In \S\,\ref{sec:breakup}, we illustrate the breakup mechanism and how a further increase in the energy of the laser pulse leads to the familiar scenario where the tin droplet deforms into a radially expanding sheet. 
To summarize, in \S\,\ref{sec:phase_diagram} we present a phase diagram, combining experiments and simulations, based on the systematic variation of both $\textrm{We}$ and pressure width $\textrm{W}$ and discuss the boundaries for the different phases.

\section{Methods}

\subsection{Experimental method}\label{sec:experimental}

We refer to \cite{kurilovich_power-law_2018} and \cite{Meijer2022_transition} for a detailed description of the experimental setup. Briefly, we use a micro-sized tin liquid jet produced from a pressurized tank placed on the top of a spherical vacuum chamber kept at a base pressure of $10^{-6}$\,mbar. The jet is fragmented into a train of equally spaced droplets, depending on the frequency of the voltage pulses applied to the nozzle\,\cite{kurilovich_power-law_2018}. In the current experiments, we produce two different droplet sizes, with diameters $D_0=50, 70\,\mu\textrm{m}$, with density $\rho=7000\,\textrm{kg}/\textrm{m}^3$, surface tension $\sigma=0.544\,\textrm{N}/\textrm{m}$ and dynamic viscosity $\mu=1.8\times10^{-3}\textrm{Pa}\textrm{s}$. The droplet interacts with a circularly polarized $1064\textrm{nm}$ Nd:YAG laser pulse, with a Gaussian temporal profile of $6\textrm{ns}$ and peak energies in the range $E_{\textrm{p}}=0.3-5\,\textrm{mJ}$. The laser is focused on the droplet as a Gaussian spot with $\sim85\,\mu$m diameter at full-width at half-maximum (FWHM); the focus is not changed throughout the experiments. To visualize the laser-droplet interaction, we use a stroboscopic imaging system, based on illuminating the droplet with a temporally and spatially incoherent green light, with a wavelength of $564\pm10\textrm{nm}$ and temporal resolution of $5\textrm{ns}$. Droplets are illuminated in the center of the chamber with this light at two different angles, $90\degree$ and $30\degree$, with respect to the laser propagation axis, imaging the droplet from side-view and nearly front-view, respectively. The corresponding shadow image is collected with a CCD camera. In order to precisely synchronize the onset of laser pulse impact and shadowgraph recording with the cameras, the laser system is triggered by a delay generator. 
The frame acquisition rate is set to be equal to the laser repetition rate at $10\,\textrm{Hz}$. Given the reproducibility of our experiments, we can trace the time evolution of the droplet after laser impact with a few nanosecond precision. This is performed by delaying the illumination pulse (SP) after the laser pulse (PP). In the following, we scan SP with time steps of $200\,\textrm{ns}$, over several capillary times.

\begin{table}
  \begin{center}
    \caption{Representative parameters of this study include droplet diameter, $D_0$, capillary time, $\tau_\textrm{c}$, laser pulse energy, $E_{\textrm{p}}$, propulsion velocity, $U$, radial expansion rate, $\dot R_0$, and dimensionless numbers like propulsion Reynolds number, $\textrm{Re}$, and propulsion Weber number, $\textrm{We}$. 
    Note that the values are indicated as approximate ranges considered in this study.} \label{table}
    \renewcommand{\arraystretch}{1.2} 
    \setlength{\tabcolsep}{8pt} 
    \small
    \resizebox{\textwidth}{!}{ 
    \begin{tabular}{llcccccccc}
      \toprule 
      \toprule
      $D_0$ ($\mu\textrm{m}$) & $\tau_\textrm{c}$ ($\mu$s) & $E_{\textrm{p}}$ (mJ) & $U$ (m/s) & $\dot R_0$ (m/s) & $\textrm{Re}$ & $\textrm{We}$ \\[1pt]
      \midrule 
       50  & 16.4 & 0.3-6 & 0.17-15  & 0.02-24 & 33-3100 & 0.02-160  \\
       70  & 27.1 & 0.2-4 & 0.20-8.7  & 0.05-10 & 54-2400 & 0.04-68  \\
       \midrule 
      \bottomrule 
      \bottomrule
    \end{tabular}
    }
  \end{center}
\end{table}

\subsection{Computational method}\label{sec:CFD}

We perform simulations to numerically predict the laser-induced deformation of a tin droplet at low Weber numbers. The governing equations for the isothermal incompressible bi-phase (droplet and ambient) flow are the continuity and momentum conservation, given by
\begin{align}
\rho\left( \frac{\partial \Ubold}{\partial t} + \nabla \cdot ( \Ubold\Ubold ) \right)	& = - \nabla p  + \nabla\cdot\left( 2 \mu \textbf{D} \right)  + \mathbf{f}_{\sigma}, \label{eq:navier1} \\
\nabla \cdot \Ubold & = 0, \label{eq:navier2}
\end{align}
where $\Ubold$ and $p$ are the velocity and pressure fields, $\textbf{D} = \left[\nabla\Ubold + (\nabla\Ubold)^T\right]/2$ is the deformation rate tensor, and $\rho$ and $\mu$ are the fluid density and viscosity, respectively. We note that, in this one-fluid formulation, $\rho$ and $\mu$ are functions with values that change across the droplet-ambient interface. The expression for these functions is given later in this paragraph. In the numerical method used here, the surface tension force is defined as a body force $\mathbf{f}_{\sigma} = \sigma \kappa \delta_s \textbf{n}$, where $\kappa$ is the local curvature of the interface, $\sigma$ the constant surface tension coefficient, $\textbf{n}$ is the unit vector normal to the interface, and $\delta_s$ is the Dirac delta function centered on the interface \cite{popinet2009accurate, Tryggvason2011-book}. The droplet interface is tracked using a volume of fluid (VOF)  scheme \citep{Hirt1981}, in which a scalar color function $c(\textbf{x}, t)$ indicates the fraction of droplet fluid contained in each numerical cell. The local density and viscosity are obtained by linearly interpolating using the value of $c$. So $\rho$ and $\mu$ from equation \eqref{eq:navier1} are defined as
\begin{align}
    \rho(c) = & \  c \ \rho_d + (1 - c)\rho_a, \label{eq:density_viscosity_dim_1} \\
    \mu(c) = & \ c \  \mu_d + (1 - c)\mu_a,
\label{eq:density_viscosity_dim_2}
\end{align}
where the indices $d$ and $a$ refer to the properties of the droplet and ambient fluids, respectively. While in experiments the droplet is contained within a vacuum chamber, due to numerical limitations, we keep the ambient fluid properties set to $\rho_a = 10^{-3} \rho_d$ and $\mu_a = 10^{-3} \mu_d$.

Equations \eqref{eq:navier1}-\eqref{eq:navier2} can be nondimensionalized by rescaling variables with the following choices:

\begin{equation}
    \textbf{x} = D_0\bar{\textbf{x}}, \hspace{10pt} t = \frac{D_0}{U}\bar{t}, \hspace{10pt} \Ubold = U\bar{\Ubold}, \hspace{10pt} p =  \rho \,U^2\bar{p}, \hspace{10pt} \kappa = \frac{1}{D_0}\bar{\kappa}, \hspace{10pt} \delta_s = \frac{1}{D_0}\bar{\delta_s},
\label{eq:nondim_scales}
\end{equation}
where $D_0$ is the diameter of the droplet, and $U$ is the droplet propulsion velocity obtained after the laser interaction.

Substituting \eqref{eq:nondim_scales} into \eqref{eq:navier1}-\eqref{eq:navier2}, we obtain the non-dimensional version of the governing equations, given by
\begin{align}
\bar{\rho}\left( \frac{\partial \bar{\Ubold}}{\partial t} + \nabla \cdot ( \bar{\Ubold}\bar{\Ubold} ) \right)	& = - \nabla \bar{p}  + \frac{1}{Re}\nabla\cdot\left( 2 \bar{\mu} \bar{\bf{D}} \right)   + \frac{1}{We} \bar{\kappa} \bar{\delta_s} \textbf{n}, \label{eq:nondim_navier1} \\
\nabla \cdot \bar{\Ubold} & = 0, \label{eq:nondim_navier2}
\end{align}
where

\begin{equation}
   Re = \frac{\rho_d \, U_z \, D_0}{\mu_d}, \quad \mathrm{and} \quad We = \frac{\rho_d \, U_z^2 \, D_0}{\sigma}
\label{eq:nondim_numbers}
\end{equation}
are the Reynolds and Weber numbers, respectively. The nondimensional density and viscosity functions are obtained from \eqref{eq:density_viscosity_dim_1}-\eqref{eq:density_viscosity_dim_2} and are given by
\begin{align}
    \bar{\rho}(c) = & \  c  + (1 - c)\rho_a/\rho_d, \\
    \bar{\mu}(c) = & \ c  + (1 - c)\mu_a/\mu_d.
\label{eq:density_viscosity}
\end{align}

Equations \eqref{eq:nondim_navier1}-\eqref{eq:nondim_navier2} are numerically solved using the open-source free software language Basilisk C~\citep{Popinet2013-Basilisk}. The droplet is created at the center of a square domain $[-5D_0, \ 5D_0] \times [-5D_0, \ 5D_0]$ that is fully discretized with a non-uniform quadtree grid~\cite{popinet2003gerris,popinet2009accurate}. To accurately resolve the flow structure inside the droplet and its shape, we apply increased refinement levels for the liquid phase and also at the interface. The maximum quadtree level of refinement used is 13, resulting in grid cells with a minimum size of $\Delta = 10D_0/(2^{13}) = 0.0012D_0$.


The volume fraction field $c$ is then advected over time by solving the equation
\begin{equation}
    \del{c}{t} + \nabla \cdot (c\, \textbf{u}) = 0.
\label{eq:vof_advection}
\end{equation}

The numerical code then solves the governing equations using a projection method and a multilevel Poisson solver. More details of the VOF implementation, including extensive numerical validation of the software language Basilisk C, can be found in many other works involving deformable surfaces, such as \cite{Sanjay2021, popinet2009accurate, Popinet2015}.

The interaction between laser pulse and droplet is modeled through the approach of \cite{gelderblom_drop_2016} as also detailed in \cite{francca2025laser}. This approach relies on the assumption that the pressure impulse experienced by the droplet occurs on a timescale much smaller than any of the relevant hydrodynamic scales. Within this very short time span, we assume that the flow inside the drop is inviscid, irrotational, and incompressible, resulting in the classical Laplace equation $(\nabla \bar{p} = 0)$ for potential flows. This equation is solved semi-analytically in spherical coordinates within the droplet by imposing the chosen pressure profile (see fig. \ref{fig:1}a,b) as a boundary condition on the drop surface. From the obtained pressure field, the potential flow assumption gives a velocity field that can be calculated through $\bar{\Ubold}_0 = -\nabla \bar{p}$. This velocity field is used as an initial condition for the full Navier-Stokes equations eqs.\,\eqref{eq:nondim_navier1}-\eqref{eq:nondim_navier2}, which are then solved in time and space according to the scheme described above.

\section{Results and discussion}\label{sec:results}
\subsection{Droplet deformation and pressure profile}\label{sec:scaling}

\begin{figure}
    \centering
    \includegraphics[width=0.75\linewidth]{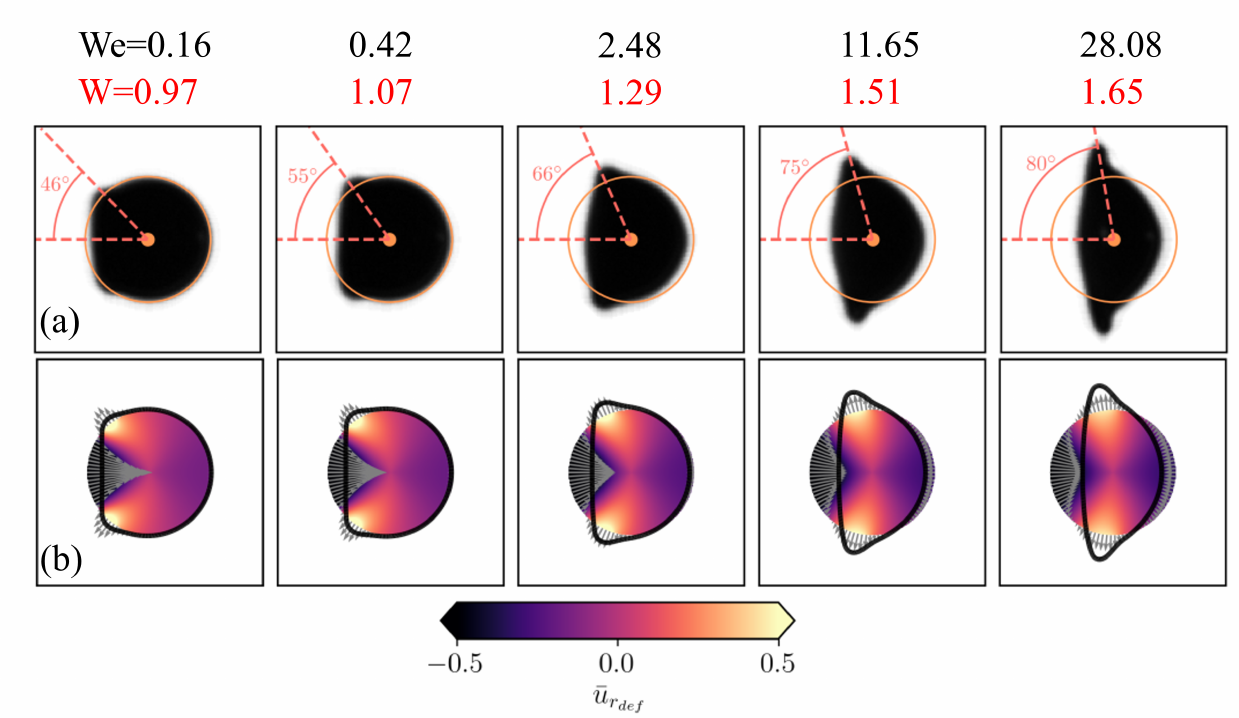}
    \caption{Droplet early deformation for different pressure profile imprinted by the laser pulse. (a) Quantification of initial droplet deformation on laser-faced side within the inertial timescale, at 200\,ns\,$\sim0.01\tau_\textrm{c}$. The impulsed liquid flow manifests as two bulges that display certain opening angles. The orange circle represents the shape of the droplet before laser impact. (b) Corresponding simulations with the same $\textrm{We}$ values with a $\textrm{W}$ parameter selected to match the angle of the surface maximum radial velocity with the corner bulges shown in (a). The gray arrows define velocity field of the liquid upon laser impact, whereas the color map shows the radial component of the velocity,  $\bar{u}_{r_{\textrm{def}}}$. The black contour represents the droplet morphology at $0.01\tau_\textrm{c}$.}
    \label{fig:2}
\end{figure}

\begin{figure}
    \centering
    \includegraphics[width=0.9\linewidth]{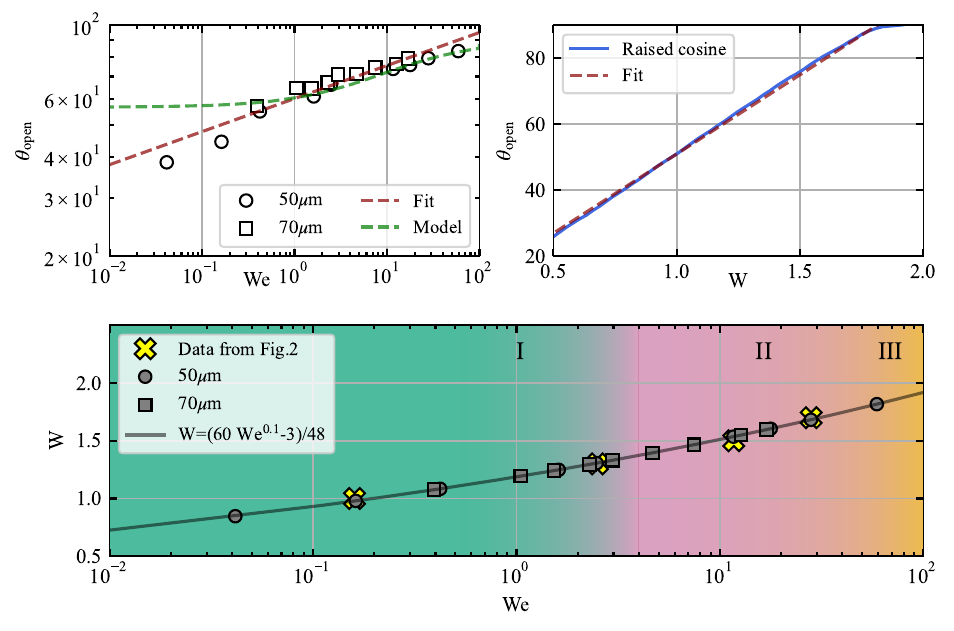}
    \caption{Correlation between the propulsion $\textrm{We}$ and the pressure width $\textrm{W}$. (a) Opening angle $\theta_{\textrm{open}}$ over $\textrm{We}$ for two droplet diameters, $D_0=50,\,70\,\mu\textrm{m}$. The red dashed line represents the empirical fit to the experimental data, with $\theta_{\textrm{open}}=60\textrm{We}^{0.1}$. The green dashed line depicts the numerical result from the model. See main text. (b) Variation of $\theta_{\textrm{open}}$ for different pressure widths $\textrm{W}$ as observed from simulations. The red dashed line depicts the numerical scaling found from simulations, with $\theta_{\textrm{open}}=48\textrm{We}+3$. (c) Variation of $\textrm{W}$ with $\textrm{We}$. The gray dots correspond to data depicted in (a) and yellow crosses show the characteristic examples shown in fig.\,\ref{fig:2}(a). Three different regimes are highlighted following experimental data: I.oscillation, II.breakup, and III.sheet formation, respectively. The gray solid line shows the correlation between $\textrm{We}$ and $\textrm{W}$ obtained from the two previous fits in (a) and (b), as illustrated by eq.\,(\ref{eq:We-vs-W}).}
    \label{fig:3}
\end{figure}

Our study requires knowledge of both propulsion $\textrm{We}$ and the width $\textrm{W}$ of the pressure profile. 
The latter is related to plasma pressure and cannot be directly obtained from experimental shadowgraphy images that track the liquid mass.
However, the fingerprint of the pressure profile  
appears as an early surface deformation on the side facing the laser, measurable at the first available frame at $200\textrm{ns}\sim0.01\tau_\textrm{c}$.
In fig.\,\ref{fig:2}(a), for $\textrm{We}=0.16$, we can see a small bulge with a certain opening angle $\theta_{\textrm{open}}\sim46\degree$. This value increases as the corresponding $\textrm{We}$ increases with increasing laser pulse energy. 
An empirical fit to the experimental data ($\theta_{\textrm{open}}$, We) yields $\theta_{\textrm{open}}\sim\textrm{We}^{0.1}$, as illustrated in fig.\,\ref{fig:3}(a). 
An immediate question arises as to whether this bulge corresponds to an impact-induced surface capillary wave traveling along the droplet. Considering the ``deep pool'' limit (\cite{lamb1905}, also see \cite{denner_dispersion_2017} or \cite{ersoy_capillary_2019}), the phase velocity of such a wave can be defined as:
\begin{equation}\label{eq:cw-velocity}
    c\approx\left(\frac{2\pi\sigma}{\rho\lambda}\right)^{1/2},
\end{equation}
where $\lambda$ is the wavelength. 
The side view of the wave suggests a wavelength of $\lambda\sim4\,\mu\textrm{m}$, with phase velocity $c\sim10\,\textrm{m/\textrm{s}}$. 
After $200\,\textrm{ns}$ the displacement of such a wave would be around $2\,\mu\textrm{m}$ which is negligible compared to the arc length observed between the upper and bottom bulges (in case of $\theta_{\textrm{open}}\sim46\degree$, this would be $\sim100\,\mu\textrm{m}$). 
This suggests that 
the observed early time deformation is a result of the plasma pressure recoil on the droplet and relates $\theta_{\textrm{\textrm{open}}}$ with W. 
This relation is straightforwardly obtained from our simulations, as illustrated in fig.\,\ref{fig:3}(b), and a linear fit to this simulation data, yielding $\theta_{\textrm{\textrm{open}}}\sim48\,\textrm{W}+3$.
Using this relation, we obtain a good agreement between experiment and simulations, shown in fig.\,\ref{fig:2}. 
A closer inspection of the snapshots in fig.\,\ref{fig:2}(b) permits visualization of the surface radial velocity, depicted by gray arrows. 
Note that the colormap within the droplet also represents this radial velocity component. 
As argued by \cite{gelderblom_drop_2016}, and \cite{francca2025laser}, the radial velocity $u_{\bar{r}_{\textrm{def}}}$ is a function of both the pressure profile and the maximum energy deposited and its maximum closely follows $\theta_{\textrm{open}}$. 
Equating the observed scalings from figs.\,\ref{fig:3}(a,b), for the current experiment we can effectively correlate $\textrm{We}$ with $\textrm{W}$ and obtain the following empirical scaling relation
 
\begin{equation}\label{eq:We-vs-W}
    \textrm{W}=\frac{\left( 60\,\textrm{We}^{0.1}-3\right)}{48}.
\end{equation}

From eq.\,(\ref{eq:We-vs-W}), we can obtain the required input on $\textrm{W}$ in the following. 
Besides the agreement between simulations and experiments at early times (cf. fig.\,\ref{fig:2}), these two parameters are sufficient to reproduce the droplet dynamics on the capillary timescale ($\sim50\,\mu\textrm{s}$), as shown in figs.\,\ref{fig:4}(a,b). 
The curve depicted in fig.\,\ref{fig:3}(c) represents the angular extension of the pressure profile on the droplet's surface (W) and its variation with $\textrm{We}$. 
Although we use an empirical fit to correlate $\textrm{We}$ with $\textrm{W}$, let us consider a simple theoretical model.  
First, note that the separate relation between propulsion (We) and laser pulse energy is well established (\cite{kurilovich_plasma_2016}, \cite{liu_mass_2020}). The energy deposited on the droplet $E_{\textrm{od}}$ is related to the pulse energy $E_{\textrm{p}}$ as $E_{\textrm{od}}=E_{\textrm{p}}\left( 1-2^{-D_0^2/d^2}\right)$, where $d=2\sigma_{\textrm{b}}$ is the diameter of the beam, with $\sigma_\textrm{b}=\textrm{FWHM}/2\sqrt{2\,\textrm{ln}2}$. 
Second, the resulting propulsion velocity $U$ of the droplet due to plasma recoil pressure scales with $E_{\textrm{od}}$ as $U=k\left(E_{\textrm{od}}-E_{\textrm{od},0}\right)^{0.6}$, where $E_{\textrm{od},0}\sim0.04\,\textrm{mJ}$ is the offset related to a threshold of plasma formation and $k=34\,\textrm{m}\,\textrm{s}^{-1}\,\textrm{mJ}^{-0.6}$ is a numerical constant following \citet{kurilovich_power-law_2018} for a $D_0=47\mu\textrm{m}$ case that is close to the current conditions. From this scaling, the dependence between the $E_{\textrm{od}}$ and propulsion based $\textrm{We}$ is obtained as 
\begin{equation}\label{eq:eod-vs-we}
    E_{\textrm{od}}=\left(\frac{\textrm{We}\,\sigma}{k^2\rho D_0}\right)^{5/6}+E_{\textrm{od},0}.
\end{equation}
We next argue that the threshold for plasma formation together with the \emph{local} laser intensity determines $\theta_{\textrm{open}}$. 
The plasma threshold can be expressed in terms of laser intensity as the ratio between the threshold laser fluence $F_{\textrm{th}}$ and pulse length: $I_{\textrm{th}}=F_{\textrm{th}}/\tau_{\textrm{p}}$ where $F_{\textrm{th}}=A^{-1}\,\rho\,\Delta H\sqrt{\mathcal{K}\,\tau_{\textrm{p}}}$ (see for example \cite{chichkov_femtosecond_1996} or \cite{aden1992laser}) with laser absorption coefficient $A$ (see below), latent heat of evaporation $\Delta H=2.5\times10^6\,\textrm{J kg}^{-1}$ and thermal diffusivity $\mathcal{K}=16.4\,\textrm{m}^2\textrm{s}^{-1}$. The fluence of the laser beam is given by the ratio between the pulse energy and the beam area $F_{\textrm{p}}=E_\textrm{p}/\pi \sigma_\textrm{b}^2$. 
The resulting beam intensity projected on the droplet's surface (see \cite{reijers2018-alignment}) can be defined as a function of $\theta$ (also see fig.\,\ref{fig:1}a)  
\begin{equation}\label{eq:projection}
    I(\theta)=\cos{\theta}\,\textrm{exp}\left[-\frac{\sin^2{\theta}}{2\alpha^2}\right], 
\end{equation}
where $\alpha=\sigma_{\textrm{b}}/R_{\textrm{eff}}$ is the dimensionless ratio between the beam width and droplet's effective radius. Here, $R_{\textrm{eff}}\gt R_0$, as a consequence of the plasma cloud formed on the laser-facing droplet pole, that effectively absorbs laser energy several micrometers away from the liquid surface. 
In our experimental range of low laser energies, close to the offset value $E_{\textrm{od,0}}$, where the plasma has not yet fully developed, it is, however, reasonable to consider a negligible plasma cloud radius and set $R_{\textrm{eff}}\approx R_0$.    
Note that in the case of $\textrm{FWHM}\gg D_0$, the eq.\,(\ref{eq:projection}) reduces to to a much simpler form $I(\theta)\sim\cos{\theta}$ enabling an analytic solution (see Appendix\,\ref{app:3}). Combining eq.\,(\ref{eq:eod-vs-we}) with eq.\,(\ref{eq:projection}), we obtain the total intensity projected on the droplet as 
\begin{equation}\label{eq:total-intensity}
    I_{\textrm{od}}(\theta)=\frac{1}{\tau_\textrm{p}\pi\sigma_\textrm{b}^2}\frac{1}{(1-2^{-D_0^2/d^2})}\left[\left(\frac{\textrm{We}\,\sigma}{k^2\rho D_0}\right)^{5/6}+E_{\textrm{od},0}\right]\cos{\theta}\,\textrm{exp}\left[-\frac{\sin^2{\theta}}{2\,\alpha^2}\right].
\end{equation}
This equation correlates $\textrm{We}$ and $\theta$ with the deposited laser intensity on droplet. If we consider that the plasma forms when the intensity on droplet overcomes the intensity threshold, $I_{\textrm{od}}(\theta)\gt I_{\textrm{th}}$, we may estimate the radial position of plasma onset for a given $\textrm{We}$ by equating $I_{\textrm{od}}(\theta_{\textrm{open}})=I_{\textrm{th}}$ to yield an estimate for $\theta_{\textrm{open}}$.
The resulting model predictions (for $D_0=50\mu\textrm{m}$) are displayed in fig.\,\ref{fig:3}(a) (also see Appendix\,\ref{app:3} for further details). We note that this simple model reproduces the experimental data overall reasonably well for the given set of the corresponding parameters and leaving $A$ as the sole free fit parameter to yield $A=0.35$. 
We note that this best fit value for $A$ is higher than the previously reported by \citet{meijer2022nanosecond} who used $A=0.16$ derived from the Fresnel equations assuming no plasma formation, which may be due to the fact that inverse bremsstrahlung absorption in the plasma increases laser absorption after plasma inception at the pole. 
The model excellently captures the experiment at higher We number, while overestimating the opening angle at the lowest We values. 
This behavior at the lowest We numbers may be expected given that eq.\,(\ref{eq:eod-vs-we}) considers only integral energy values, not a local fluence, and the opening angle at low We is sensitive to $E_{\textrm{od},0}$ (see \S\,\ref{app:3} for further details). 
Furthermore, \citet{kurilovich_power-law_2018} also pointed out difficulty explaining the onset behavior in eq.\,(\ref{eq:eod-vs-we}) at low We even when employing full radiation hydrodynamics modeling. 
All in all, our simple model captures the essence of the behavior: W increases monotonically with increasing We. 
In the following, we mainly focus on the influence of We and W on droplet deformation without further consideration of the underlying driving plasma dynamics. To this end, we use the empirical fit eq.\,(\ref{eq:We-vs-W}) that is fundamental for understanding the complete picture of droplet deformation driven by low-energy laser pulses.
Furthermore, from eq.\,(\ref{eq:We-vs-W}) we learn that these two parameters are inherently correlated, as an increase in $\textrm{We}$, implies a wider region of overlap for the plasma-induced pressure on the droplet. 
Finally, based on a qualitative inspection of our experimental data shown in fig.\,\ref{fig:3}(c), we classify the observed dynamical regimes of the droplet, namely, capillary-driven oscillations, inertia-induced breakup, and sheet formation (fig.\,\ref{fig:1}(c-e)), highlighted by the colormap.
In the following sections, we will explain these behaviors in detail. 

\subsection{Droplet oscillation and surface waves}\label{sec:oscillation}

\begin{figure}
    \centering
    \includegraphics[width=1\linewidth]{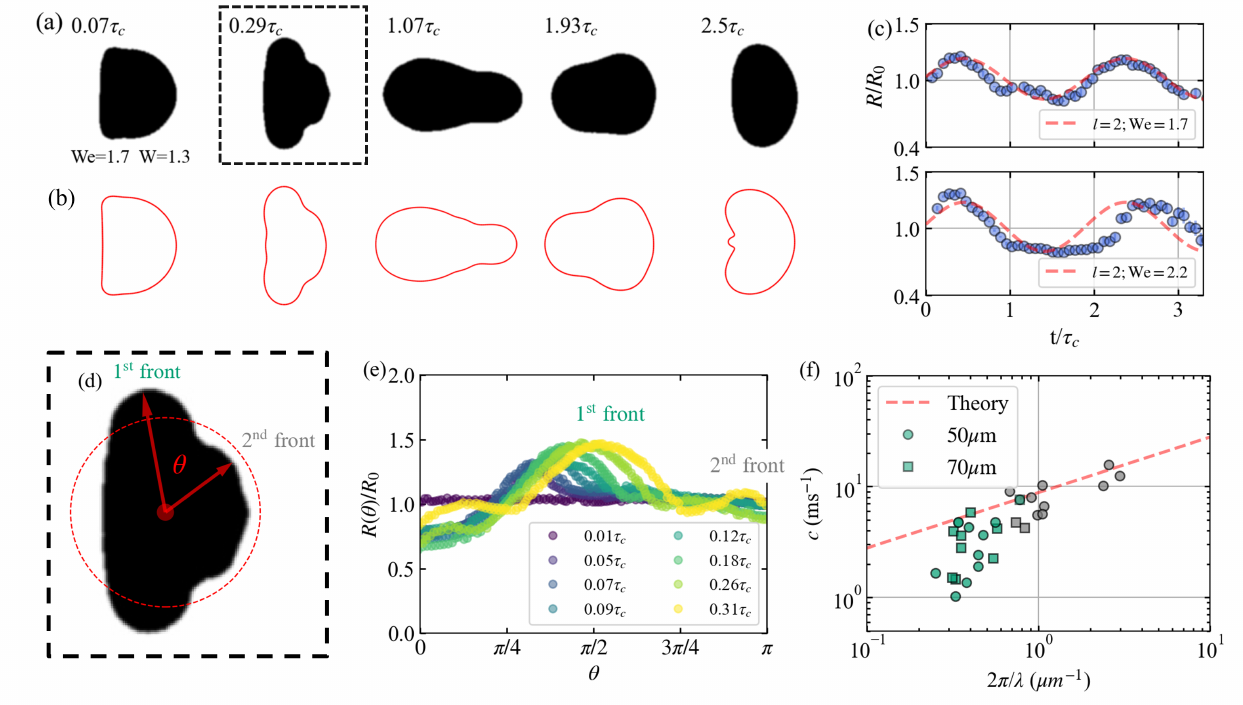}
    \caption{Droplet oscillation. (a) Shadowgraphs of droplet deformation within the oscillation regime at different fractions of capillary time $\tau_\textrm{c}$ for $\textrm{We}=1.7$ and $\textrm{W}=1.3$. (b) Numerical results at the same times as depicted in (a) and the same values for $\textrm{We}$ and $\textrm{W}$. (c) Nondimensional radius $R/R_0$ over time $t/\tau_\textrm{c}$ for two different oscillation cases: $\textrm{We}=1.7$ and $\textrm{W}=1.3$ (upper plot), $\textrm{We}=2.2$ and $\textrm{W}=1.4$ (bottom plot). The red dashed line corresponds to the best fit of the oscillation estimated from the equation of Rayleigh modes with $l=2$. (d) Example of a staircase-like structure where surface capillary waves (CW) are pointed out with red arrows as ``$1^{\textrm{st}}$ and $2^{\textrm{nd}}$ fronts''. The red dashed circle represents the shape of the droplet at rest. Here, $\theta$ is the radial position of the CW front on the surface. (e) Parametric representation of the surface contour for different times $t/\tau_\textrm{c}$ as nondimensional radial extension $R(\theta)/R_0$ over $\theta$. The two CW fronts can be observed as two peaks. Data includes droplet at rest (straight line at $0.01\tau_\textrm{c}$) and at several instances after impact to illustrate the origin and propagation of CW. (f) CW phase estimated for $1^{\textrm{st}}$ (green data) and $2^{\textrm{nd}}$ (blue data) fronts. The red line represents the dispersion law depicted by eq.\,(\ref{eq:cw-velocity}).}
    \label{fig:4}
\end{figure}

Starting from low-energy pulses, we first observe the droplet oscillation as described in \S\,\ref{sec:intro} and depicted in fig.\,\ref{fig:4}(a) where we show an example of oscillation at $\textrm{We}=1.7$ and associated $\textrm{W}=1.3$ [cf. eq.\,(\ref{eq:We-vs-W})]. 
In fig.\,\ref{fig:4}(b), we display the numerical results at the same instants with the same values of $\textrm{We}$ and $\textrm{W}$ as in (a). Note the good agreement between the experimental data and the simulation. In figs.\,\ref{fig:4}(a,b), after the characteristic initial deformation, explained in the previous section, the droplet expands radially (see frames at $0.29\tau_\textrm{c}$). Here, the radial expansion rate is much lower than the propulsion velocity, i.e., $\dot R_0\ll U$, therefore, capillary retraction overcomes any inertia-driven upward flow. Consequently, the droplet retracts and oscillates. If we trace the radial extension over time, as shown in the upper plot of fig.\,\ref{fig:4}(c), plotting the non-dimensionalized vertical half size, $R(\theta$=$\pi/2)/R_0$, over time, we clearly distinguish the first cycle of oscillation. Despite the complexity of shapes observed over time in fig.\,\ref{fig:4}(a), the overall temporal dynamics is surprisingly well described by a small-amplitude capillary-driven body oscillation, the so-called Rayleigh mode, with oscillation frequency $\omega_\textrm{R}^2=\sigma l\left( l-1\right)\left(l+2\right)/(\rho R_0^3)$, where $l=2$ as the fundamental mode for axisymmetric expansion/contraction. Rayleigh modes have also been studied in the context of free-flying droplets\,\cite{khismatullin_shape_2001,bostwick_capillary_2009_oscillations}, pendant droplets\,\cite{basaran1994nonlinear_pendant,moon2006lowest_pendant_oscillation}, or droplets flowing in different liquid media\,\cite{abi_chebel_shape_2012}. In this particular case, for a droplet with $D_0=50\,\mu\textrm{m}$, the oscillation period, $t_\textrm{R}=2\pi/\omega_\textrm{R}\approx27\,\mu\textrm{s}$. The assumption of small amplitude oscillation appears reasonable given the maximum height within the first cycle, $\approx0.2D_0=12\,\mu\textrm{m}\lt D_0=50\,\mu\textrm{m}$. Increasing $\textrm{We}$ will make more liquid flow in the radial direction, resulting in stronger retraction and higher amplitude of oscillation. Indeed, we can see such an increase in amplitude in the bottom plot of fig.\,\ref{fig:4}(c) for $\textrm{We}=2.0$. Moreover, in the bottom plot of fig.\,\ref{fig:4}(c) we notice that the overall oscillation dynamics is no longer described by the Rayleigh mode with $l=2$. Instead, the curve quickly deviates from the theory, showing a long valley between $t\sim0.5\,\tau_\textrm{c}$ to $2.5\,\tau_\textrm{c}$. This valley corresponds to a longer horizontal expansion of the droplet. 



A closer inspection of the droplet during its initial deformation reveals a staircase-like structure, particularly evident in fig.\ref{fig:4}(a,d) at $t=0.29\,\tau_\textrm{c}$. 
This structure consists of successive capillary wavefronts triggered by laser impact, shaping the droplet as waves propagate. 
Similar patterns have been observed in droplet impact on solids (\cite{renardy_pyramidal_2003}, \cite{li_capillary_2019}). 
Here, the laser pulse initiates capillary waves (CW), driving liquid accumulation into a bulge that grows and propagates, forming subsequent wavefronts. 
Red arrows in fig.\,\ref{fig:4}(d) mark the various wavefronts. 
Interfacial contour profiles at different times, as shown in fig.\ref{fig:4}(e), illustrate the phase dynamics. The first CW front increases in amplitude and propagates azimuthally. Around $t=0.31\,\tau_\textrm{c}$, a second front appears with a significantly lower amplitude. In some cases, for a larger droplet ($D_0=70\,\mu\textrm{m}$), a third CW front is also observed.  
After estimating the wavelengths $\lambda_{\textrm{CW}}$ from peak full-width at half-maximum (FWHM) in fig.\,\ref{fig:4}(e), we plot phase velocity $c$ versus wavenumber $k=2\pi/\lambda_{\textrm{CW}}$ in fig.\,\ref{fig:4}(f), comparing experimental data for $D_0=50, 70\,\mu\textrm{m}$ with eq.\,(\ref{eq:cw-velocity}). 
Green dots represent the first CW front, deviating from eq.\,(\ref{eq:cw-velocity}) as may be expected from its large amplitude ($R(\theta)/R_0\sim1.5$), invalidating the "deep pool" assumption. 
In contrast, the second and third fronts, with lower amplitudes ($\lesssim 1.1R_0$), behave as surface capillary waves, closely following the predicted velocity. 
In conclusion, the oscillating droplet shows a complex combination of bulk oscillation and surface waves.

\subsection{Droplet breakup}\label{sec:breakup}

\begin{figure}
    \centering
    \includegraphics[width=0.75\linewidth]{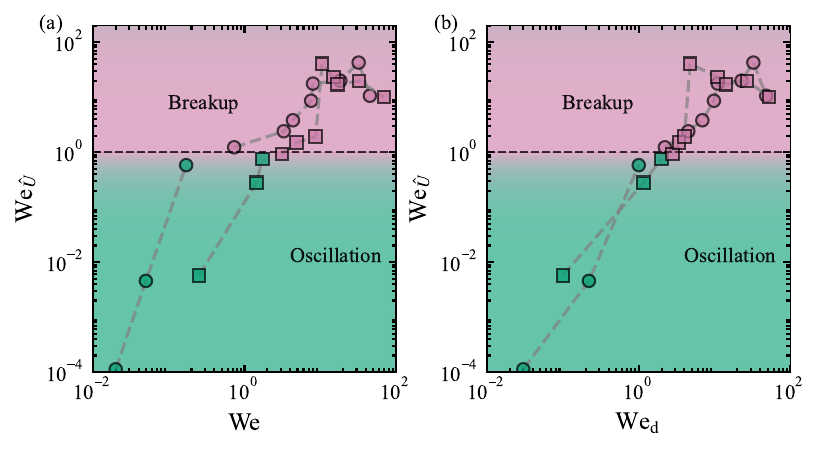}
    \caption{Droplet breakup. (a) Variation of the critical Weber number $\textrm{We}_{\hat{U}}$ based on the horizontal expansion rate $\hat{U}$ after radial retraction over different propulsion Weber numbers $\textrm{We}$, for two different droplet diameters, $D_0=50, 70\,\mu\textrm{m}$ (circle and square symbols, respectively). (b) Data from panel (a) with the horizontal axis represented by deformation Weber number $\textrm{We}_\textrm{d}$. The limit between oscillation and breakup regimes is estimated at $\textrm{We}_{\hat{U}}=1$, as shown with the dashed line, with droplet oscillation for $\textrm{We}_{\hat{U}}\lt1$ and droplet breakup for $\textrm{We}_{\hat{U}}\gt1$.}
    \label{fig:5}
\end{figure}

Droplets break up at large oscillation amplitudes. 
Hereafter, we refer to the horizontally stretched droplet as a filament, which expands horizontally at a certain velocity $\hat{U}$. 
Since the expansion of this filament is balanced by the kinetic energy of accumulating mass within the tip and the surface tension pulling back the filament, we can define the corresponding Weber number as $\textrm{We}_{\hat{U}}=\rho \hat{U}^2 D_f/\sigma$, where $D_f$ is the characteristic diameter of the filament before the breakup. 
Figure\,\ref{fig:5}(a) shows a monotonic increase of $\textrm{We}_{\hat{U}}$ with $\textrm{We}$, with the filament containing more mass that is moving at larger velocities, with increasing We. 
This behavior is well captured in fig.\,\ref{fig:5}(a) with the data points corresponding to two different droplet diameters following the same trend. 
Based on the balance between the inertia and capillary forces acting on the filament, we expect that the dominance of the tip inertia over capillary retraction will lead to the filament breakup. 
Conversely, if the surface tension overcomes the horizontal extension of the filament, the droplet will oscillate. 
A qualitative visualization of the experimental data supports this argument, as we observe droplet oscillation for $\textrm{We}_{\hat{U}}\lt1$, and droplet breakup for $\textrm{We}_{\hat{U}}\gt1$ for both droplet sizes independently.  
This observation closely aligns with previous studies of inertia-driven breakup in liquid threads, insofar as they define a critical $\textrm{We}$ number for breakup (see \cite{clanet1999transition}). 
Our critical Weber number $\textrm{We}_{\hat{U}}$ clearly captures the transition into breakup. 
However, the propulsion velocity-based impact $\textrm{We}$ alone does not dictate the breakup behavior or determine $\textrm{We}_{\hat{U}}$. 
A highly focused pressure impulse (small $W$) may result in violent breakup even in the absence of significant propulsion. 
Indeed, we also observe different curves for different droplet diameters ($D_0=50,70\,\mu\textrm{m}$, circles and squares) when using We as the sole driving parameter. 
Instead, since the droplet oscillation/breakup is caused by the relative radial expansion and retraction, we invoke a more suitable parameter that is the \emph{deformation }Weber number based on radial expansion rate, $\dot{R}_0$, as $\textrm{We}_\textrm{d}=\rho \dot{R}_0^2 D_0/\sigma$. 
Following \cite{liu_speed_2022}, we define $\dot{R}_0$ to be the velocity orthogonal to $U$: it captures the vertical expansion of the liquid.  
In fig.\,\ref{fig:5}(b) we rescale the horizontal axis with $\textrm{We}_\textrm{d}$ and capture the variation of $\textrm{We}_{\hat{U}}$ that is now nearly invariant to the droplet diameter. 
Finally, we can observe that at much higher $\textrm{We}_\textrm{d}$ values, the filament velocity gradually decreases. 
The slowing of the filament extension is due to the fact that, at higher $\textrm{We}$, even though the radial flow is stronger, the retraction becomes less effective as a sheet begins to develop.

\subsection{Sheet formation}\label{sheet-formation}

The dynamics of a radially expanding sheet following droplet impact at high We numbers (We $\gtrsim$100) is well studied (see e.g. \cite{villermaux_drop_2011}, \cite{wang_bourouiba_2018_rim}) and we omit a detailed characterization of this regime in the current work. 
\cite{klein_drop_2020} and \cite{liu_speed_2022} demonstrated that, following a laser-droplet impact, the sheet expansion rate is accurately captured by the deformation Weber number, $\textrm{We}_\textrm{d}$, as defined in the preceding section. 
Indeed, the choice for $\textrm{We}_\textrm{d}$ in Sec.\,\ref{sec:breakup} was inspired by these works. 
The change in radius over time is described by balancing radial acceleration and surface tension continuously pulling back the sheet, resulting in the formation of a bounding thick rim (\cite{wang_bourouiba_2018_rim}), which is a source of ligaments and fragments that drain the sheet on the capillary timescale. 
\cite{liu_speed_2022} illustrated that the time-varying radius of the sheet scales with $\textrm{We}_\textrm{d}$ as $R(t)/R_0\sim\textrm{We}_\textrm{d}^{1/2}f\left( t/\tau_\textrm{c}\right)$, for considerably higher range of values for $\textrm{We}_\textrm{d} \sim 1000-10\,000$. 
The polynomial function $f\left( t/\tau_\textrm{c}\right)$ depends on the flow rate of the liquid from the sheet into the rim (see \cite{wang_bourouiba_2017_thickness} for details). 
This is due to the required radial acceleration that initiates the formation of the rim. 
According to \cite{wang_bourouiba_2018_rim}, this happens when local instantaneous capillary length equals the rim diameter, thus, resulting in the local Bond number $\textrm{Bo}=\rho(-\ddot{R})b/\sigma=1$, where $\ddot{R}$ and $b$ are sheet radial acceleration and rim thickness, respectively. 
This condition is fulfilled when the radial inertia overcomes the capillary retraction. In Appendix\,\ref{app-sheet}, we present sample experimental data for sheet formation at $\textrm{We}_\textrm{d}\sim300-1000$. 
For this study, we consider a sheet to be formed when the maximum radial extension of the liquid deformation exceeds twice the initial droplet radius, i.e., $R_{\textrm{max}} > 2R_0$. 
For the laser-droplet impact case, it is known that the maximum radius follows the relation $R_{\textrm{max}}-R_0 \sim0.14\sqrt{\textrm{We}_\textrm{d}}R_0$ (see, e.g., \cite{liu_speed_2022}, and see \cite{villermaux_drop_2011}, \cite{wang_bourouiba_2018_rim} for the droplet-pillar case) which indicates that the criterion for sheet expansion is met for $\textrm{We}_\textrm{d}=60$. 
It is this criterion that establishes the second and last boundary in the phase map that is presented next.

\subsection{Phase map}\label{sec:phase_diagram}

\begin{figure}
    \centering
    \includegraphics[width=0.8\linewidth]{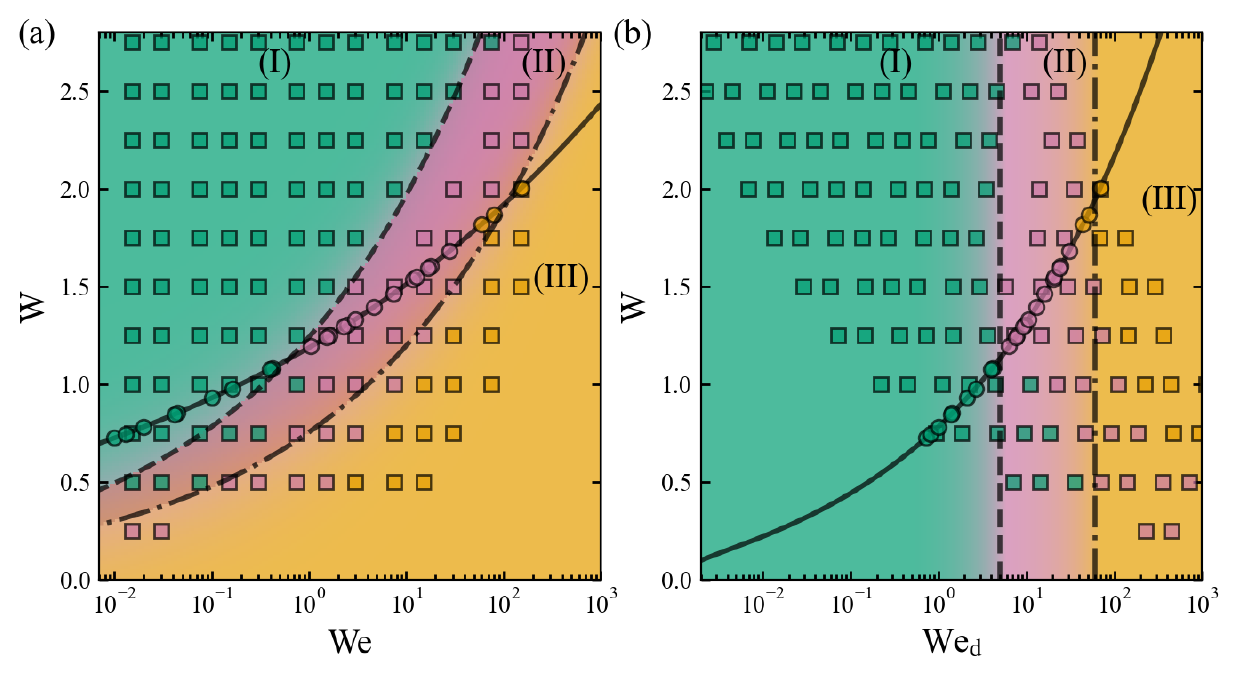}
    \caption{(a) Phase diagram of droplet dynamics as a function of center mass propulsion $\textrm{We}$ number and pressure width, $\textrm{W}$. Circles correspond to experimental data, while squares represent the simulations. Three different regimes are identified: oscillation (I), breakup (II), and sheet expansion (II). The black solid line corresponds to eq.\,(\ref{eq:We-vs-W}). Examples of each regime is illustrated in figs.\,\ref{fig:1}(c-e). (b) The same phase diagram but representing $\textrm{W}$ as the function of $\textrm{We}_\textrm{d}$. Black dashed and dot-dashed lines in panels (a,b) represent the scaling $\textrm{W}\sim\left(\textrm{We}/\textrm{We}_\textrm{d}\right)^{1/5}$ with $\textrm{We}_\textrm{d}=5$ and $60$, as limits between oscillation/breakup and breakup/sheet formation regimes, respectively (see the main text). Note that these vertical lines are shown as curves in (a).
    }
    \label{fig:6}
\end{figure}

Using the criteria outlined above, the droplet behavior analyzed in previous sections can be portrayed in a single phase diagram based on the two key parameters: the pressure width $\textrm{W}$ and impact $\textrm{We}$ number.  
As discussed in \S\,\ref{sec:scaling}, these two parameters are inherently correlated in the current experiments.
However, via numerical computations we can build a two-dimensional map with $\textrm{We}$ and $\textrm{W}$ as independent parameters.
In Fig\,\ref{fig:6}(a) we illustrate a combination of experimental data (circles) and simulations (squares) for a wide range of both parameters, and we define different regimes based on specific criteria. 
The corresponding scaling from eq.\,(\ref{eq:We-vs-W}) is depicted with a black line that necessarily goes through the experimental data points.  
We observe three well-defined regimes, corresponding to droplet oscillation [green area, (I)], droplet breakup [red area, (II)], and sheet formation [orange area, (III)]. 
Recall that the transition between regimes (I) and (II) is straightforwardly identified in both experiment and simulations (breakup onset).  
To distinguish droplet breakup from sheet formation, we focus on the droplets maximum radial expansion. 
Recall that we choose the sheet formation onset as $R_{\textrm{max}}=2R_0$.  
Considering previous studies in aerodynamic breakup of liquid droplets (see \cite{guildenbecher2009secondary}), it is reasonable to invoke radial flow as an additional parameter to account for the underlying mechanism. 
Moreover, we have already discussed the relevance of radial expansion both for breakup onset (see section \S\,\ref{sec:breakup}) and sheet formation. 
From our experiments (supported by simulations) we observe breakup onset at $\textrm{We}_\textrm{d}\sim5$. Furthermore, the criteria $R_{\textrm{max}}=2R_0$ is fulfilled at $\textrm{We}_\textrm{d}\sim60$. 
Similar values were found in previous studies (see \cite{guildenbecher2009secondary}, \cite{jackiw2021aerodynamic}) for droplet breakup and sheet thinning behaviors. 
To obtain the explicit dependence of $\textrm{W}$ with $\textrm{We}$ and $\textrm{We}_\textrm{d}$, we recall our previous studies where these two nondimensional numbers are related as $\textrm{We}_\textrm{d}\sim (\dot{R}_0/U)^2 \textrm{We}$ (see \cite{hernandez-rueda_early-time_2022}). 
Also, the term $(\dot{R}_0/U)^2$ strongly depends on the angular projection of the plasma generated upon laser impact. 
In the particular case of the raised cosine beam profile, we simulated the dependence of $\dot{R}_0/U$ with $\textrm{W}$ and determined an empirical scaling $\dot{R}_0/U\sim A\textrm{W}^{-5/2}$, with constant of proportionality $A\sim3.8$ (also see the Appendix of \cite{francca2025laser}). 
If we insert this equation into the relation between two Weber numbers, we obtain $\textrm{We}_\textrm{d}\sim A^2\textrm{W}^{-5}\textrm{We}$, which permits relating the pressure width $\textrm{W}$ with propulsion $\textrm{We}$ as $\textrm{W}\sim\left(\textrm{We}/\textrm{We}_\textrm{d}\right)^{1/5}$. 
This allows us to represent the phase diagram in terms of $\textrm{W}$ and $\textrm{We}_\textrm{d}$ as shown in fig.\,\ref{fig:6}(b). 
Effectively, we observe the limits as vertical lines, revealing that the transition between regimes is well captured by just the deformation Weber number, $\textrm{We}_\textrm{d}$. 
We note that the definition of $\textrm{We}_\textrm{d}$ in the simulations, following \cite{francca2025laser}, slightly differs from the experiment: the simulations use the maximum velocity in the co-moving frame (which can be conveniently extracted already from the initialization of the velocity field), which is not necessarily along the vertical $r$ axis in the $r-z$ cylindrical coordinate frame cf. fig.\,\ref{fig:1}.  
Any differences between the two definitions are minor, except for the cases with the smallest values of W (at large We) where the simulations overestimate $\textrm{We}_\textrm{d}$. 
Indeed, this difference in definition is partially responsible for the discrepancies in the phase diagram at low W values. 
The resulting phase diagram shows that oscillations occur for $\textrm{We}\textrm{d} < 5$, the droplet breaks up for $5 \le \textrm{We}\textrm{d} \le 60$, and a radially expanding sheet forms for $\textrm{We}\textrm{d} > 60$, consistent with the condition $R_{\textrm{max}}=2R_0$. 
Given these two values of $\textrm{We}_\textrm{d}$, the corresponding phase diagram can be represented in terms of $\textrm{W}$ and $\textrm{We}$, where, based on the identified scaling, the vertical lines in fig.\,\ref{fig:6}(b) become curves in fig.\,\ref{fig:6}(a), following $\textrm{W} \sim \textrm{We}^{1/5}$. 

\section{Conclusions}\label{sec:conclusions}

In this work, we analyze the droplet dynamics upon nanosecond laser impact at low Weber numbers ($0.1-100$). Below the threshold of sheet formation, the droplet displays a complex interplay between radial flow and center mass propulsion, leading to different deformation behaviors. 
We identify three regimes: droplet oscillation, breakup, and finally, sheet formation. 
After laser impact, the droplet radial expansion is followed by capillary-mediated retraction, resulting in axisymmetric oscillations. 
Droplet oscillations overall are found to follow the small-amplitude body motion characterized by Rayleigh modes. 
Superimposed on the body oscillations are capillary waves. 
We predominantly observe fundamental Rayleigh mode (expansion/retraction), with higher amplitudes as $\textrm{We}$ is increased. 
Eventually, this leads to droplet breakup when the horizontal deformation is long enough. 
A critical Weber number $\textrm{We}_{\hat{U}}$ is introduced that captures the transition to breakup. 
We find that the whole dynamics is captured as a balance between the inertia of the horizontal motion and surface tension. 
Furthermore, we identify the role of the recoil pressure profile's radial distribution on the surface of the droplet and demonstrate that the higher the laser peak energy, resulting in higher $\textrm{We}$, the wider the pressure width, denoted by $\textrm{W}$. 
We show the intrinsic correlation between these two parameters that is present in the experiment. 
Complemented with simulations, we propose a phase diagram based on $\textrm{We}$ and $\textrm{W}$, and study the droplet deformation as solely determined by pressure width, propulsion velocity, and radial deformation rate, captured by $\textrm{We}_\textrm{d}$ as the single parameter defining the phase diagram.  

Our studies closely correlate with industrial applications for nanolithography, where the source of extreme ultraviolet radiation is based on the interaction of nanosecond laser pulses with micro-sized liquid tin droplets. 
We extend the knowledge of previous studies on the topic toward the uncharted regime of low-energy laser pulse interacting with tin droplets. 
Furthermore, we highlight the role of surface radial distribution of the pressure profile and its correlation with the propulsion of the droplet. 
This relation is not evident from the case of the sheet formation, as it always comprises extreme $\textrm{We}$ values which, given the aforementioned correlation, come with a large $\textrm{W}$. 

Furthermore, the results of the present study not only deepen our understanding of low-amplitude droplet deformations, but also directly relate to other applications of laser-liquid interactions. For instance, in Laser-Induced Forward Transfer (LIFT) \cite{serra2019laser,jalaal2019laser,das2024review}, where precise control of plasma formation near fluid and solid interfaces controls the jet formation \cite{jalaal2019destructive} or in Laser Ablation in Liquids \cite{yang2007laser,yan2012pulsed}, where plasma ablates the solid target into the surrounding fluid. 
Nonetheless, many questions remain open. The detailed dynamics of laser-induced plasma at the liquid surface demand further investigations, since this ultra-fast multi-scale process crucially influence the fluid mechanism of the problem. More directly related to the present work, examining the complex mode composition within an oscillating droplet prior to breakup and understanding the transition from symmetric splitting to filament jetting are natural next steps.

\begin{acknowledgments}
This work was conducted at the Advanced Research Center for Nanolithography (ARCNL), a public-private partnership between the University of Amsterdam (UvA), Vrije Universiteit Amsterdam (VU), Rijksuniversiteit Groningen (UG), the Dutch Research Council (NWO), and the semiconductor equipment manufacturer ASML and was partly financed by ‘Toeslag voor Topconsortia voor Kennis en Innovatie (TKI)’ from the Dutch Ministry of Economic Affairs and Climate Policy. The authors were supported by funding from the European Research Council (ERC) under the European Union’s Horizon 2020 research and innovation programme under grant agreement No 802648, and the OTP grant with project number 19458 financed by the Dutch Research Council (NWO). 
\end{acknowledgments}

\section*{Data Availability Statement}
The data that support the findings of this study are available from the corresponding author upon reasonable request.

\section*{Competing interests}

The authors are not aware of any conflict of interest that might affect the objectivity of this study.

\section*{Author ORCID}

\noindent M. Kharbedia, \url{https://orcid.org/0000-0002-2128-9945}

\noindent H. Fran\c ca, \url{https://orcid.org/0000-0002-5361-7704}

\noindent H. K. Schubert, \url{https://orcid.org/0009-0000-4499-0091}

\noindent D.J. Engels, \url{https://orcid.org/0000-0001-7363-8716}

\noindent M. Jalaal, \url{https://orcid.org/0000-0002-5654-8505}

\noindent O. Versolato, \url{https://orcid.org/0000-0003-3852-5227}

\newpage

\appendix

\section{Sheet formation}\label{app-sheet}

\begin{figure}
    \centering
    \includegraphics[width=0.8\linewidth]{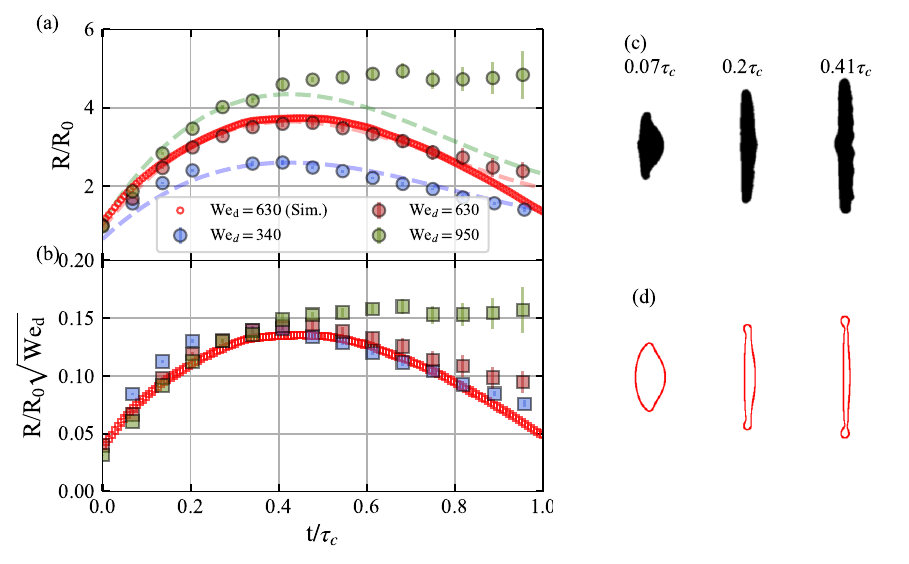}
    \caption{Radially expanding sheet formation. 
    (a) Variation of the dimensionless sheet radius, $R/R_0$ with time $t/\tau_\textrm{c}$ for four different $\textrm{We}_\textrm{d}$ that includes experimental data: $\textrm{We}_\textrm{d}=340,\,670,\,950$, and simulation: $\textrm{We}_\textrm{d}=670$. 
    The dashed lines represent eq.\,(\ref{eq:sheet-expansion}). 
    (b) The collapse of the curves represented in (a) by rescaling the vertical axis with $R/R_0\sqrt{\textrm{We}_\textrm{d}}$. (c) Side-view frames of expanding thin sheet at three different instances depicted as fractions of capillary time for $\textrm{We}_\textrm{d}$, and (d) Frames obtained from simulations for $\textrm{We}_\textrm{d}=670$.}  
    \label{app:2}
\end{figure}

Upon high-energy laser impact, the droplet deforms radially with $\dot{R}_0\sim U$ (see the main text).  
The expanding dynamics is determined by the formation of the bounding liquid rim. 
As shown in previous studies for laser-droplet interaction (see for example \cite{klein_drop_2020}, \cite{liu_speed_2022}), radial extension is described by balancing the inertia of the rim and surface tension of the sheet. 
Following our studies on tin sheets, we describe the variation of the sheet radius through
\begin{equation}\label{eq:sheet-expansion}
    \frac{2R}{D_0\sqrt{\textrm{We}_\textrm{d}}}=A\left(t/\tau_\textrm{c}\right)^3+B\left(t/\tau_\textrm{c}\right)^2+C\left(t/\tau_\textrm{c}\right)+D,
\end{equation}
where the fit parameters are $A=0.33(5)$, $B=-0.92(5)$, $C=0.57(2)$ and $D=0.035(2)$. 
The numbers within brackets are standard deviation uncertainty. In fig.\,\ref{app:2} we present expansion curves for $\textrm{We}_\textrm{d}=340, 670, 950$. For comparison, we include simulated data for $\textrm{We}_\textrm{d}=670$.
We can see a qualitatively good match between eq.\,(\ref{eq:sheet-expansion}) and the experimental data in fig.\,\ref{app:2}(a). 
In the particular case of $\textrm{We}_\textrm{d}=950$, the sheet breaks before retraction, leading to a noticeable deviation from the theoretical curve at later times. 
In fig.\,\ref{app:2}(b), we collapse the data from panel (a) onto a single curve by rescaling dimensionless radius with $\textrm{We}_\textrm{d}$, revealing the universal behavior of the sheet. Finally, to illustrate the similarity between experimental observation and numerical simulations, in figs.\,\ref{app:2}(c,d) we display frames at three different times. 

\section{Plasma onset and droplet deformation}\label{app:3}

\begin{figure}
    \centering
    \includegraphics[width=0.8\linewidth]{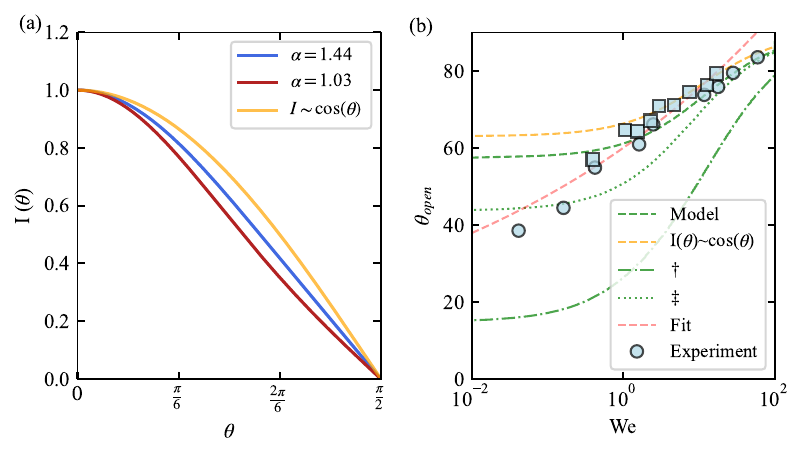}
    \caption{(a) Projection of the laser intensity on a spherical surface. The intensity $I(\theta)$ variation over $\theta$ on the laser-facing droplet pole ($0-\pi/2$), for different laser width to droplet radius ratios: $\alpha=1.44$ for $R_0=25\,\mu\textrm{m}$ and $\alpha=1.03$ for $R_0=35\,\mu\textrm{m}$. The orange line represents the intensity projection as approximated by a cosine function $I(\theta)\sim\cos{\theta}$. (b) Variation of the opening angle $\theta_{\textrm{open}}$ with $\textrm{We}$. Blue circles ($R_0=25\,\mu\textrm{m}$) and squares ($R_0=35\,\mu\textrm{m}$) depict experimental observations as shown in fig.\,\ref{fig:3}(a). The green dashed line is the result of the numerical computation of eq.\,(\ref{eq:total-intensity}). The red dashed line represents the empirical fit depicted by eq.\,(\ref{eq:We-vs-W}). The orange dashed line represents the analytical solution for $\theta$ in eq.\,(\ref{eq:total-intensity}), when approximated by a cosine function ($\alpha \gg 1$ case). Additionally, we show two more solutions for eq.\,(\ref{eq:total-intensity}): with values $A=0.15$ and $E_\textrm{od,0}=0.04\,\textrm{mJ}$ ($\dag$) and with values $A=0.35$ and $E_\textrm{od,0}=0.025\,\textrm{mJ}$ (see appendix text for discussion)}.
    \label{fig:9}
\end{figure}

Here we further detail the model developed in \S\,\ref{sec:scaling}. We assume that the opening angle $\theta_{\textrm{open}}$ observed in fig.\,\ref{fig:2}(a) corresponds to the largest angle $\theta$ where the plasma is still generated given a certain distribution of the projected intensity on droplet $I_{\textrm{od}}$ together with a threshold intensity $I_{\textrm{th}}$. 
The intensity distribution is described by eq.\,(\ref{eq:total-intensity}) for a given value of propulsion based $\textrm{We}$. Equation\,(\ref{eq:total-intensity}) is solved numerically by equating $I_{\textrm{od}}=I_{\textrm{th}}$. 
We compute eq.\,(\ref{eq:total-intensity}) for $\textrm{FWHM}=85\,\mu\textrm{m}$ and $D_0=50\,\mu\textrm{m}$. The non-dimensional ratio between beam width and droplet's effective radius is defined as $\alpha=\sigma/R_{\textrm{eff}}$. In the limit of small energy on droplet values (close to the offset value $E_{\textrm{ed,0}}$), we consider $R_{\textrm{eff}}\approx R_0$. We choose the absorption coefficient $A$ as a (sole) free fit parameter given that beyond but near plasma threshold the laser absorptivity $A$ will range between the value given by the Fresnel equations for an undisturbed liquid ($A\approx0.15$ for unpolarized light) and that given by a fully developed plasma ($A\approx1$). In fig.\,\ref{fig:9}(b) we show the result of a fit to the data leading to $A=0.35$ a value indeed in the expected range.  
For the sake of comparison, in fig.\,\ref{fig:9}(a) we illustrate the intensity projection for different values of $\alpha$. 
Note that the projection can be approximated by a cosine function for $\alpha\gg1$ in which case it is possible to obtain an analytical solution to eq.\,(\ref{eq:total-intensity}), resulting in $\theta\sim\textrm{arccos}(\textrm{We}^{6/5})$. This solution is shown in fig.\,\ref{fig:9}(b) as an orange dashed line (taking $A=0.35$ for this also). Since the range of laser fluences used for low $\textrm{We}$ number dynamics is close to the threshold for plasma onset, it is not possible to elucidate the exact dynamics of plasma formation. We note that neither the numerical solution nor the cosine approximation fully accounts for the experimental observations in the range of $\textrm{We}$ values. Notably, the higher deviations are observed at low values $\textrm{We}\lt0.1$. As explained in the main text, this regime might be characterized by a not fully developed plasma, giving rise to a highly nontrivial interplay between pressure profile and the projected laser intensity. Additionally, to clarify the influence of the choice of $E_\textrm{od,0}$, we next compute eq.\,(\ref{eq:total-intensity}) for the same absorption coefficient $A=0.35$ but using the offset value $E_\textrm{od,0}$ as the free parameter to match the lower regime. The resulting curve is depicted in fig.\,\ref{fig:9}(b) ($\ddag$) with $E_\textrm{od,0}=0.025\textrm{mJ}$. Note that this solution implies a lower plasma threshold than previously considered to describe the overall effect of plasma pressure leading to propulsion.  
It may well be that the threshold value found previously for the propulsion threshold (see e.g. \citet{kurilovich_power-law_2018}), a result of integrated overall plasma pressure, does not represent well the \emph{local} plasma dynamics. 
Indeed, the curve resulting from eq.\,(\ref{eq:total-intensity}) for these input values (see fig.\,\ref{fig:9}(b), $\dag$) undershoots the experimental data.  
Furthermore, as noted by \citet{kurilovich_power-law_2018} no simple relation exists that accurately translates laser intensity to local plasma pressure. 


\newpage



\end{document}